\documentclass[aps,pra,amsmath,amssymb,twocolumn,groupedaddress]{revtex4-2}
\usepackage[english]{babel}

\usepackage{amsmath}
\usepackage{amssymb}
\usepackage{graphicx}
\usepackage{stfloats}
\usepackage{lipsum}
\usepackage{physics}
\usepackage{braket}  
\usepackage[version=4]{mhchem}
\usepackage{bbold}
\usepackage{xcolor}
\usepackage{float}
\usepackage{longtable}
\usepackage{orcidlink}
\usepackage{comment}

\hypersetup{
  colorlinks=true,
  linkcolor=black,
  citecolor=black,
  urlcolor=black
}
\AtBeginDocument{%
}
\usepackage{graphicx}
\usepackage{epstopdf}  
\setcitestyle{square,numbers}   
\usepackage{natbib}
\renewcommand{\thesection}{\Roman{section}.}
\renewcommand{\thesubsection}{\Alph{subsection}.}

\makeatletter
\renewcommand{\p@subsection}{\thesection}
\renewcommand{\@seccntformat}[1]{%
  \ifcsname format@#1\endcsname
    \csname format@#1\endcsname
  \else
    \csname the#1\endcsname\quad
  \fi
}
\newcommand{\format@subsection}{ ~\thesubsection}
\makeatother

\begin{document}

\title{Relativistic Effects on Photoabsorption Cross Sections of Highly Charged Ions}

\author{Anvar Khujakulov\orcidlink{0009-0006-9721-8533}}
 \email{anvar.khujakulov@uni-jena.de}
\affiliation{Institut f\"ur Festk\"orpertheorie und Optik, Friedrich-Schiller-Universit\"at Jena, 07743 Jena, Germany}

\author{Caterina Cocchi\orcidlink{0000-0002-9243-9461}}
\affiliation{Institut f\"ur Festk\"orpertheorie und Optik, Friedrich-Schiller-Universit\"at Jena, 07743 Jena, Germany}
\affiliation{Abbe Center of Photonics, Friedrich-Schiller-Universit\"at Jena, 07745, Jena, Germany}

\begin{abstract}
The study of highly charged ions offers a unique platform for probing the breakdown of non-relativistic theory under the influence of extreme electromagnetic environments. Here, we investigate the photoabsorption of highly charged ions within the dipole approximation using both the time-dependent Schr\"odinger equation (TDSE) and the time-dependent Dirac equation (TDDE), modelling the external field as an instantaneous broadband excitation. Nonrelativistic scaling relations with respect to the nuclear charge are utilized as a diagnostic tool to systematically identify and quantify relativistic contributions. Within the purely nonrelativistic TDSE framework, these scaling relations hold exactly, allowing the absorption spectra of arbitrary highly charged ions to be inferred directly from a neutral hydrogenic reference. However, as the nuclear charge increases, relativistic effects become dominant through a sizeable blue shift in the absorption cross section, due to the relativistic enhancement of the binding energy. We further evaluate semi-relativistic TDSE approximations by direct comparison with full TDDE simulations, assessing their predictive power and establishing the regimes where a full Dirac treatment is indispensable for quantitative accuracy.

\end{abstract}

\maketitle
\section{Introduction}\label{sec.intro}
Recent advances in extreme light sources, such as the European X-ray Free-Electron Laser (XFEL) \cite{sfa:xfel17_1} and the Linear Coherent Light Source (LCLS) \cite{gen:christ16}, have opened a new frontier of light-matter interaction, providing access to radiation intensities on the order of $10^{23}~\mathrm{W/cm^2}$ or beyond. These modern facilities combine unprecedented peak brightness with attosecond-scale pulse durations~\cite{duris2020tunable,yan2024terawatt}, enabling the coherent excitation and real-time tracking of electronic wavepackets in the X-ray spectral range~\cite{maroju2023attosecond,guo2024experimental}.
In this ultra-intense regime, the electron dynamics become inherently non-perturbative and relativistic, as the quiver velocity of liberated electrons approaches the speed of light \cite{sfa:reis92,sfa:reiss14}. The states generated under these extreme conditions are currently the subject of pioneering experimental research at large-scale facilities~\cite{sfa:spil06,sfa:guen08,gen:gsi14}. Recently, Son et al.~\cite{sfa:Son2025} highlighted the complexities of interpreting XFEL-driven spectra, demonstrating that resonance-like features in multi-resonant X-ray scattering can arise from sequential absorption rather than true resonances. Complementary theoretical work by M\"uller et al.~\cite{sfa:Mueller2024} has further established the necessity for high-precision Dirac-based modeling, achieving sub-meV agreement with experimental measurements of double-K-hole resonances in He-like ions. These advances reveal the complexity of accurately deciphering the spectroscopic signatures of highly charged ions, where transient polarization and high-frequency coherence dominate the early-time dynamics.

From a theoretical perspective, a rigorous characterization of the electron dynamics at relativistic velocities requires the solution of the time-dependent Dirac equation (TDDE). 
While this formalism intrinsically incorporates relativistic kinematics and spinorial corrections, such as spin-orbit coupling and the Darwin term \cite{sfa:rath97,sfa:sels09,sfa:vann12,sfa:torKj17}, it lacks the elegant universal scaling symmetry found in the time-dependent Schr\"odinger equation (TDSE). In the non-relativistic dipole approximation, any hydrogen-like ion can be mapped exactly onto a hydrogen atom benchmark through a space-time coordinate transformation. This scaling has served as a cornerstone for energy calibration and cross-section predictions across the periodic table \cite{sfa:shah87,sfa:mads99c,sfa:Bray20}. However, as the nuclear charge $Z$ increases, this universal symmetry relation inevitably breaks.

The central goal of this work is to utilize scaling relations as a physical diagnostic tool for relativistic states. By comparing full TDDE solutions against scaled TDSE benchmarks, we systematically isolate and quantify purely relativistic spectral signatures. To achieve this goal, we employ a theoretical idealization: perturbing the system with a broadband, instantaneous $\delta$-kick excitation. Following the formalism introduced by Yabana and Bertsch~\cite{quad:Yabana1996} for real-time time-dependent density-functional theory, this approach impulsively populates the entire excited-state manifold~\cite{quad:cocchi21}, providing the response of the system to the full spectral range. While the $\delta$-kick method has recently become a valuable tool for describing optical nonlinearities~\cite{quad:cocchi21,quad:Cocchi14,quad:Driouech2025}, we focus here on the weak-field regime to isolate the fundamental influence of relativistic kinematics on the spectral blue shift. This analysis provides the necessary baseline for validating numerical frameworks before moving toward the highly complex regimes of strong-field multiphoton excitation. Within this framework, we evaluate the predictive power of semi-relativistic approximations, including effective nuclear charge ($Z'$) mapping \cite{sfa:vann12} and field-dressed mass modifications \cite{sfa:tork18,sfa:ivano18}, to determine where these shortcuts succeed and where full Dirac dynamics are indispensable.

The article is organized as follows: In Sec.\ref{sec.rels}, we summarize the theoretical framework and the implementation of the length-gauge solvers, designed to effectively suppress negative-energy states in the relativistic regime \cite{sfa:klai06,sfa:band13}. After defining the numerical settings in Sec.~\ref{sec:computational},  we report our results in Sec.~\ref{sec.rels1}, where we demonstrate the emergence of the relativistic blue shift and the breakdown of non-relativistic scaling for highly charged ions. Finally, we provide a comparative analysis of semi-relativistic mapping techniques and summarize our conclusions in Sec.~\ref{sec:conclu}.

\section{Theoretical framework}\label{sec.rels}

\subsection{Numerical methods for solving TDSE and TDDE}\label{ssec.num_SE.DE}

\subsubsection{The non-relativistic approach}\label{ss.nonreTh}
We model the interaction between a hydrogen-like ion and the external field as a broadband instantaneous excitation, represented by a $\delta$-kick \cite{quad:Yabana1996}, arbitrarily oriented along the $z$-axis:
\begin{equation}\label{eq.delF}
\mathbf{F}(t) = \hat{z}F_0\delta(t),
\end{equation}
where $F_0$ denotes the field amplitude. Such an impulsive excitation provides a uniform, infinitely broadband spectral distribution that populates the entire excited-state manifold simultaneously. This idealization allows the intrinsic spectroscopic signal, including both bound-to-bound and bound-to-continuum transitions, to be extracted from the subsequent temporal evolution of the dipole moment in a single simulation. In a laboratory context, the pulse always possesses a finite bandwidth, which acts as a frequency-domain windowing function, introducing a high-frequency cutoff (e.g., with a Gaussian profile) that suppresses the excitation of high-lying states near the ionization threshold. While this spectral filtering modifies the initial wavepacket composition and dampens the absolute amplitude of the transient polarization, the underlying relativistic physics investigated here remains qualitatively robust.

The electron dynamics initiated by the broadband excitation are governed by the TDSE in the length-gauge dipole approximation \cite{quad:Yabana1996}:
\begin{equation}\label{eq.tdse1}
i\hbar \frac{\partial\psi_{\rm nr}(\textbf{r},t)}{\partial t}=
\left[ \hat{H}_{\mathrm{nr},0} + F_0\delta(t)z \right] \psi_{\rm nr}(\textbf{r},t),
\end{equation}
where
\begin{equation}
    \hat{H}_{\mathrm{nr},0} = -\frac{\hbar^2}{2\mu}\nabla^2_r -\frac{Ze^2}{4\pi\varepsilon_0 r}
\end{equation}
represents the field-free Hamiltonian and $\mu$ is the reduced mass. The subscript ''nr'' denotes non-relativistic operators and wavefunctions. In Eq.~\eqref{eq.tdse1}, the interaction term $F_0\delta(t)z$ acts as an impulsive driving force, imparting an instantaneous phase shift $e^{-iF_0 z/\hbar}$ onto the initial state at $t=0^+$.

For the numerical solution of the TDSE [Eq.\eqref{eq.tdse1}], we exploit the universal scaling symmetry of the non-relativistic dipole Hamiltonian. We expand the time-dependent wavefunction $\psi_{\rm nr}(\mathbf{r},t)$ into a complete basis formed by the eigenstates $\phi(\mathbf{r})$ of the field-free nonrelativistic Hamiltonian $\rm \hat{H}_{\mathrm{nr},0}$~\cite{sfa:coll89,sfa:vann12}:
\begin{equation}\label{eq.pSI_F}
\psi_{\rm nr}(\textbf{r},t)=\sum_{j}c_{j} e^{-iE_{\mathrm{nr},j}t/\hbar}\,\phi_{j}(\mathbf{r}) \quad (t > 0),
\end{equation}
where time-independent expansion coefficients $c_{j}$ are determined by the $\delta$-kick at $t=0$. The sum runs over all discretized eigenstates (including both bound and continuum states), with $j \equiv \{n_j,\ell_j,m_j\}$ encoding the complete set of quantum numbers. The eigenstates of the field-free Hamiltonian,
\begin{equation}\label{eq.ds3}
\hat{\rm H}_{\mathrm{nr},0}(\mathbf{r})\, \phi(\mathbf{r}) = E_{\mathrm{nr}}\, \phi(\mathbf{r}),
\end{equation}
are factorized into radial functions and spherical harmonics,
\begin{equation}\label{eq.radan}
\phi(\mathbf{r}) = \frac{\mathcal{R}_{n,\ell}(r)}{r} \cdot Y_{\ell, m}(\theta, \phi),
\end{equation}
with the radial function represented using a B-spline basis set \cite{bsp:boor78}:
\begin{equation}\label{eq.bspline}
{\cal R}(r)=\sum^{n}_{i=1}p_{i}B_{i+1}(r).
\end{equation}
The radial coordinate is confined to $r \in [0, R_{\mathrm{max}}]$ with the first and the last spline being omitted to enforce the boundary conditions: ${\cal R}(r=0)={\cal R}(r=R_{\rm max})=0$. Substituting Eq.~\eqref{eq.radan} into Eq.~\eqref{eq.ds3} yields a generalized eigenvalue problem of size $n\times n \; \forall \; \ell$, which is diagonalized using the \texttt{LAPACK} routine \texttt{DSBGVX}. 

In the $\delta$-kick formalism~\cite{quad:Yabana1996}, the initial non-equilibrium state is expressed by:
\begin{equation}\label{eq.psi1}
\psi_{\rm nr}(\textbf{r},0^+) = e^{-izF_0/\hbar}\, \psi_{\rm nr}(\textbf{r},0^-).
\end{equation}
In the weak-field limit ($F_0\ll1$), the exponential part of Eq.\eqref{eq.psi1} can be expanded and rewritten in the first order as:
\begin{equation}\label{eq.psi1a}
\psi_{\rm nr}(\textbf{r},0^+) \approx \left(1 - \frac{izF_0}{\hbar}\right) \psi_{\rm nr}(\textbf{r},0^-).
\end{equation}
Taking the initial state $\psi_{\rm nr}(\textbf{r},0^-)$ as the field-free ground state $\phi_0(\mathbf{r})$, the first-order correction $\delta\psi = -izF_0\phi_0(\mathbf{r})$ represents the dipole-coupled perturbation that populates the excited-state manifold via dipole transitions, determining the time-independent expansion coefficients $c_j$ in Eq.~\eqref{eq.pSI_F}.

The time-dependent dipole moment induced by the kick is evaluated as the expectation value of the $\hat{z}$ operator \cite{quad:cocchi21,sfa:feli21}:
\begin{equation}\label{eq.diP_sig_a}
{d}_{\rm nr}(t)=\langle \psi_{\rm nr}(\textbf{r},t) | \hat{z} | \psi_{\rm nr}(\textbf{r},t) \rangle.
\end{equation}
Substituting the wavefunction expansion [Eq.~\eqref{eq.pSI_F}] into Eq.~\eqref{eq.diP_sig_a} yields:
\begin{equation}\label{eq.diP_sig_b}
d_{\mathrm{nr}}(t) = \sum_{i} \sum_{j}c_i^{*} c_j \, e^{-i (E_j - E_i) t / \hbar} \, F_0 \langle \phi_i |\hat{z} | \phi_j \rangle,
\end{equation}
where the time dependence is governed by the transition frequencies $\omega_{ij} = (E_j - E_i)/\hbar$. The matrix elements $\langle \phi_i | \hat{z} | \phi_j \rangle$ are evaluated using the radial-angular decomposition [Eq.\eqref{eq.radan}] and the B-spline representation [Eq.~\eqref{eq.bspline}]:
\begin{equation}\label{eq.vij_factor}
\langle \phi_i | \hat{z} | \phi_j \rangle  = W_{ij} \, M^L_{ij}.
\end{equation}
Here, the angular factor $W_{ij}$ encodes the dipole selection rules: 
\begin{equation}\label{eq.wfi}
\begin{split}
W_{ij} = (-1)^{l_j - m_j} &\sqrt{(2l_j + 1)(2l_i + 1)} \\
&\times \begin{pmatrix}
l_j & 1 & l_i \\
-m_j & 0 & m_i
\end{pmatrix}
\begin{pmatrix}
l_j & 1 & l_i \\
0 & 0 & 0
\end{pmatrix}
\end{split}
\end{equation}
and the radial matrix element 
\begin{equation}\label{eq.mij}
M^L_{ij} = \int_0^{R_{\mathrm{max}}} dr \, r \, \mathcal{R}_i(r) \mathcal{R}_j(r)
\end{equation}
is evaluated via Gaussian quadrature up to a convergence-tested radius $R_{\mathrm{max}}$, which effectively approximates the infinite limit of the analytical treatment.

The photoabsorption cross section is computed as the Fourier transform of the dipole moment  \cite{quad:Yabana1996}:
\begin{eqnarray}\label{eq.nonsigma}
\sigma_{\rm nr}(\omega) = \frac{4\pi\omega}{c}\frac{\Im\{d_{\rm nr}^{*}(\omega)F^{*}(\omega)\}}{|F^{*}(\omega)|^2},\quad\quad
\end{eqnarray}
where $F^{*}(\omega)$ is the Fourier transform of the $\delta$-kick in Eq.\eqref{eq.delF}, yielding a constant spectrum, and $c$ is the speed of light.

\subsubsection{The relativistic approach}\label{ss.reTh}

Relativistic effects become increasingly prominent with rising nuclear charge, particularly for ions with \(Z \geq 20\) \cite{gen:gran70,gen:gran07}. In this scenario, the electron dynamics are ruled by the TDDE:
\begin{equation}\label{eq.tdde1}
i\hbar \frac{\partial\psi_{\rm r}(\textbf{r},t)}{\partial t}= \left(\hat{H}_{\mathrm{r},0}+ F_0\delta(t)z\right)\psi_{\rm r}(\textbf{r},t),
\end{equation}
where $\psi_{\rm r}(\mathbf{r},t)$ is a four-component spinor wavefunction and the interaction term is again described by a $\delta$-kick excitation. The subscript 'r' denotes relativistic quantities. In Eq.~\eqref{eq.tdde1}, the field-free Dirac Hamiltonian reads:
\begin{equation}
\hat{H}_{\mathrm{r},0} = c\boldsymbol{\alpha}\cdot\mathbf{\hat{p}} + \beta m_ec^2 - \frac{Ze^2}{4\pi\varepsilon_0 r},
\end{equation}
where $\boldsymbol{\alpha} = (\alpha_x, \alpha_y, \alpha_z)$ and $\beta$ are the four Dirac matrices satisfying the Clifford algebra $\{\alpha_i, \alpha_j\} = 2\delta_{ij}I_4$, and $\{\alpha_i, \beta\} = 0$ \cite{gen:gran07}.

While this work focuses on hydrogen-like ions, the underlying numerical methods provide a scalable framework. Specifically, the two-center Dirac equation is solved by extending an adapted B-spline basis set method originally developed for relativistic atomic calculations, a robust approach that avoids the spurious states commonly encountered in alternative methods, such as those used for $\mathrm{H}_2^+$ \cite{sfa:Zapata2024}. The broader applicability of this methodology has been reported for the ground-state energies for Bi–Au, U–Pb, and Cf–U quasimolecules~\cite{sfa:Kotov2022}, calculated within the two-center approach along with evaluations of the leading quantum electrodynamic (QED) contributions in the monopole approximation, in good agreement with prior calculations. These examples underscore the versatility of the present numerical framework and highlight the need for further development of two-center QED methods for precise quasimolecular spectra. 

From the diagonalization of the time-independent Dirac equation,
\begin{equation}\label{eq.Deig}
\hat{H}_{\mathrm{r},0}\Phi(\textbf{r})=E_{r}\Phi(\textbf{r}),
\end{equation}
using a B-spline basis expansion~\cite{sfa:vann12}, the eigenstates are represented in spinorial form as~\cite{gen:gran07}:
\begin{equation}\label{eq.dir_psi}
\Phi_{\rm \kappa m}(\textbf{r}) = \frac{1}{r} 
\begin{pmatrix}
P_{\kappa}(r) \, \chi_{\kappa, m}(\hat{\textbf{r}})\\[6pt]
i Q_{\kappa}(r) \, \chi_{-\kappa, m}(\hat{\textbf{r}})
\end{pmatrix},
\end{equation}
where $P_{\kappa}(r)$ and $Q_{\kappa}(r)$ denote the large and small radial components, respectively. The angular part is described by the $\ell s$-coupled spherical spinor $\chi_{\kappa, m}$, while the symmetry of the spinor is governed by the relativistic angular quantum number:
\begin{equation}\label{eq.kappa_def}
\kappa = \begin{cases}
\displaystyle -\left(j + \frac{1}{2}\right) = -(\ell + 1), & \text{for } j = \ell + \frac{1}{2},\\[10pt]
\displaystyle \phantom{-}\left(j + \frac{1}{2}\right) = \ell, & \text{for } j = \ell - \frac{1}{2}.
\end{cases}
\end{equation}
By substituting Eq.~\eqref{eq.dir_psi} into Eq.~\eqref{eq.Deig} \cite{gen:gran07,sfa:vann12}, the field-free Dirac eigenvalue 
problem turns into coupled radial equations for $P_{\kappa}(r)$ and $Q_{\kappa}(r)$, which are expanded in a B-spline basis 
\begin{eqnarray}\label{eq.pqD}
P(r)=\sum_{i=1}^{n}p_iB_{i+1}, \quad Q(r)=\sum_{i=1}^{n}q_iB_{i+1},
\end{eqnarray}
in analogy with the nonrelativistic case [Eq.~\eqref{eq.bspline}]. 
Inserting these expansions into the radial Dirac equations yields a $2n \times 2n$ generalized eigenvalue problem. The same diagonalization procedure used in the nonrelativistic case yields the relativistic energy spectrum and the corresponding eigenstates.

The solution of the TDDE [Eq.~\eqref{eq.tdde1}] for $t > 0$ is obtained by expanding the time-dependent wavefunction $\psi_{\rm r}(\mathbf{r},t)$ in the basis of field-free Dirac eigenstates $\Phi_{j}(\mathbf{r})$:
\begin{equation}\label{eq.pSI_Fr}
\psi_{\rm r}(\mathbf{r},t) = \sum_{j} C_{j} e^{-i E_{\mathrm{r},j} t/\hbar} \Phi_{j}(\mathbf{r}),
\end{equation}
where $j \equiv \{n, \kappa, m\}$ denotes the set of relativistic quantum numbers. Immediately following the $\delta$-kick, the wavefunction undergoes the instantaneous phase shift. Similar to the non-relativistic case  [Eq.~\eqref{eq.psi1}], this shift represents an impulsive momentum transfer to the initial spinor state, populating the relativistic excited-state manifold across the entire spectral range. In the weak-field limit, the expansion coefficients $C_j$ are determined by projecting the ``kicked'' initial state onto the relativistic basis for a given field strength $F_0$.
The relativistic dipole moment after the pulse is found in analogy with Eq.~\eqref{eq.diP_sig_a}, using the relativistic wavefunction [Eq.~\eqref{eq.pSI_Fr}]. 
The relativistic analog of Eq.~\eqref{eq.diP_sig_b} is therefore~\cite{sfa:Zapa21}:
\begin{equation}\label{eq.reldi}
d_{\rm r}(t) = \sum_{i,j} C^*_i C_j e^{-i(E{\mathrm{r},j} - E_{\mathrm{r},i})t/\hbar} \langle \Phi_i | \hat{z} | \Phi_j \rangle.
\end{equation}

For spinor states, $\Phi_i$ and $\Phi_j$, with quantum numbers $i \equiv \{n_i, \kappa_i, j_i, l_i, m_i\}$ and $j \equiv \{n_j, \kappa_j, j_j, l_j, m_j\}$, respectively, the time-independent matrix elements are computed as:
\begin{equation}
\langle \Phi_i | \hat{z} | \Phi_j \rangle = \delta_{m_j, m_i} \, \delta_{|l_j - l_i|, 1} \, W_{ji} \, M_{ji},
\end{equation}
where the angular factor $W_{ji}$ is given by:
\begin{eqnarray}
W_{ji} = (-1)^{j_j - m_j} (-1)^{j_i + 1/2} \sqrt{(2j_j + 1)(2j_i + 1)}\nonumber
\\
\times \begin{pmatrix}
j_j & 1 & j_i \\
-m_j & 0 & m_i
\end{pmatrix}
\begin{pmatrix}
j_j & 1 & j_i \\
-1/2 & 0 & 1/2
\end{pmatrix},
\end{eqnarray}
and the radial matrix element $M_{ji}$ is:
\begin{equation}
M_{ji} = \int_0^{R_{\rm max}} dr \, r \, \big[ P_i(r)P_j(r) + Q_i(r)Q_j(r) \big].
\end{equation}
After determining the relativistic dipole moment from Eq.\eqref{eq.reldi}, the relativistic photoabsorption cross-section is calculated as:
\begin{equation}\label{eq.nonsigmr}
\sigma_{\rm r}(\omega) = \frac{4\pi\omega}{c} \frac{\Im{d_{\rm r}(\omega)}}{F_0}.
\end{equation}

It is worth noting that working in the length-gauge [Eq.\eqref{eq.tdde1}] ensures that the interaction term remains diagonal in the spinor space. In the weak-field regime, this effectively decouples the positive- and negative-energy solutions, allowing the time propagation to be physically confined to the electronic subspace. This approach is consistent with the no-pair approximation~\cite{PhysRevA.32.756} and avoids the numerical and physical complications associated with the explicit evolution of positronic components. Furthermore, while observables are fundamentally gauge-invariant in an exact, untruncated Hilbert space, gauge choices heavily impact convergence in numerical implementations \cite{sfa:baue05b,sfa:vann09}. In the relativistic domain, the velocity gauge strongly couples electronic bound states to the negative-energy continuum, which requires significantly larger basis sets and explicit handling of electron-positron pairs to suppress numerical artifacts. By contrast, the length gauge effectively suppresses these negative-energy states in the weak-field regime, maintaining a robust electronic subspace and optimizing numerical convergence \cite{sfa:klai06,sfa:vann12,sfa:malt23}. Finally, since the electric field strengths employed remain several orders of magnitude below the Schwinger limit ($F \ll F_{\rm S} = m_e^2 c^3 / e \hbar \approx 2.6\cdot 10^6$~a.u.)~\cite{gen:schwi51}, any contributions from vacuum polarization or electron-positron pair production are strictly negligible.

\subsection{Exact scaling relations for the TDSE }\label{sec.theor_1}
The TDSE scaling relations serve as a benchmark for relativistic effects: any deviation from these exact mappings in full TDDE simulations represents a direct, quantifiable measure of relativistic symmetry-breaking. In the non-relativistic dipole approximation, the TDSE for a hydrogen-like ion can be mapped exactly onto the hydrogen atom reference due to the underlying $1/r$ potential symmetry \cite{sfa:mads99c}. Through a coordinate and time scaling transformation, the dynamics of an ion with nuclear charge $Z$ are analogous to those of the hydrogen atom.  For the coordinates, time, and frequency, these relations establish a theoretical benchmark:
\begin{subequations}\label{eq.scal1}
\begin{align}
r' &= \frac{\mu}{m_{e}} Zr, \\
t' &= \frac{\mu}{m_{e}} \frac{t}{Z^{2}}, \\
\omega' &= \frac{m_{e}}{\mu} \omega Z^{2},
\end{align}
\end{subequations}
where the primed variables refer to the hydrogen atom ($Z=1$) and $\mu \approx m_e$ is assumed.
For a continuous laser field with amplitude $F(t)$, the scaling is $F' = F / Z^3$. However, when using a $\delta$-kick of the form $F(t) = F_0 \delta(t)$, the transformation property of the Dirac $\delta$-function, $\delta(at) = \delta(t)/|a|$, must be taken into account. With the time scaling as $t' = Z^2 t$, the $\delta$-kick strength $F_0$ scales as:
\begin{equation}\label{eq.scal2}
F_0 = Z F'_0.
\end{equation}
The other scaling relations in Eqs.(\ref{eq.scal1}a-c) remain unchanged. Moreover, although Eq.~\eqref{eq.psi1a} provides an approximate analytical form of the wavefunction following the kick, the scaling transformations for time and pulse characteristics [Eq.~\eqref{eq.scal2}] maintain their validity when applied to the exact wavefunction [Eq.\eqref{eq.psi1}], as these scalings derive from the fundamental symmetries of the underlying equations.

%


The induced dipole moment and the photoabsorption cross section are subject to scaling relations, too. For hydrogen-like ions, they are obtained by substituting Eqs.~(\ref{eq.scal1}a-c) and Eq.\eqref{eq.scal2} into Eqs.\eqref{eq.diP_sig_a} and \eqref{eq.nonsigma}. The resulting quantities exhibit an explicit functional dependence on $Z$:
\begin{subequations}\label{eq.sigma_scale}
\begin{align}
d(t) &=\frac{1}{Z}\, d'(t'), \\
\sigma(\omega) &= \sigma'(\omega').
\end{align}
\end{subequations}
From these results, the absorption cross-section of any hydrogen-like ion can be predicted based on the known
results for the hydrogen atom by scaling the field strength according to Eq.~\eqref{eq.scal2}. 

\subsection{Semi-relativistic scaling relations for the TDSE and the TDDE }\label{sec.theor_2}
\begin{figure}[h!]
 \includegraphics[width=0.49\textwidth]{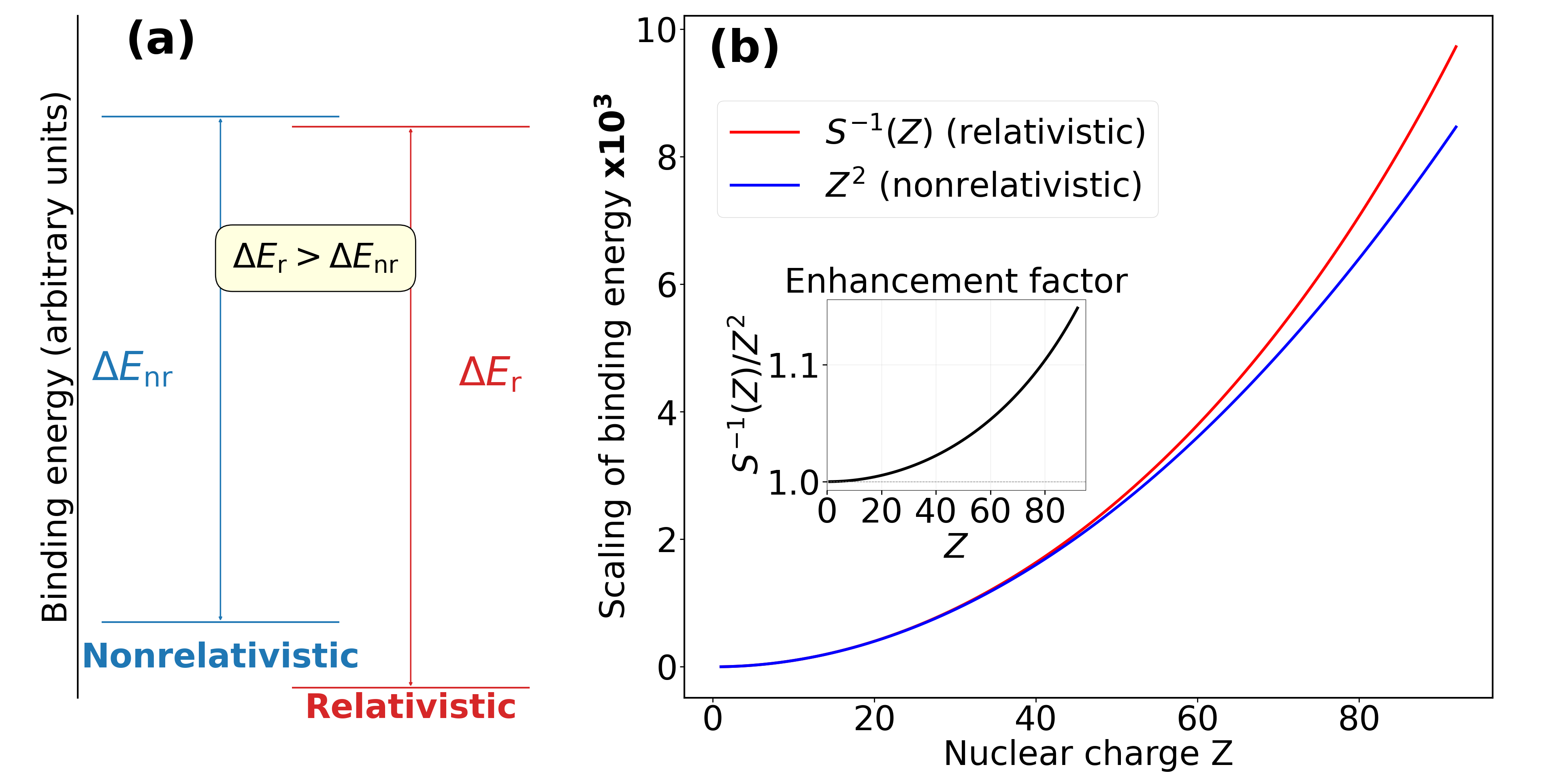}
 \caption{ (a) Schematic of $n=1 \to 2$ transitions for nonrelativistic (blue) and relativistic (red) systems. The relativistic contraction of levels increases the transition energy ($\Delta E_{\rm r} > \Delta E_{\rm nr}$). (b) Scaling of binding energies relative to $Z=1$ as a function of nuclear charge $Z$. The relativistic factor $S^{-1}(Z)$ (red) grows faster than the nonrelativistic $Z^{2}$ scaling (blue). Inset: The ratio $S^{-1}(Z)/Z^{2}$ highlights the relativistic enhancement as $Z$ increases. }
  \label{fig:dip1_a0}
\end{figure}

Semi-relativistic approaches provide a critical bridge between the non-relativistic framework of the TDSE and the fully relativistic TDDE, for which exact scaling relations are unknown. For any nuclear charge $Z$, the relativistic binding energy always exceeds its non-relativistic counterpart, as sketched in Fig.~\ref{fig:dip1_a0}a. To isolate the impact of this energy shift, we introduce an effective nuclear charge,
\begin{eqnarray}\label{eq.shz}
Z^\prime=\sqrt{2c^{2}\left(1-\sqrt{1-\left(\frac{Z}{c}\right)^{2}}\right)},\end{eqnarray}
to allow a non-relativistic electron to mimic the relativistic contraction and increased binding energy of the Dirac $1s$ ground state.
By matching the relativistic binding energy of a system with charge $Z$ to a non-relativistic model with charge $Z'$, we can evaluate the predictive power of the semi-relativistic approximations. This mapping, hereafter referred to as TDSE($Z'$), allows us to distinguish whether the observed spectral shifts arise purely from the relativistic enhancement of the binding energy or if they necessitate the full spinorial treatment of the Dirac framework \cite{sfa:vann12}.

Alternative semi-relativistic scaling relations can be derived by combining the nonrelativistic laws specified in Sec.\ref{sec.theor_1} with the relativistic energy shift \cite{sfa:ivano18}.
For a hydrogen-like ion with nuclear charge $Z$ and the hydrogen atom ($Z=1$), the electric field frequency and strength are mapped according to:
\begin{subequations}\label{eq.irina}
\begin{align}
\omega' &=\omega\,\frac{1 - \sqrt{1 - 1/c^2}}{1 - \sqrt{1 - Z^2/c^2}}:=\omega S(Z),\\[6pt]
F'_0     &=F_0\,S^{3/2}(Z),
\end{align}
\end{subequations}
respectively. Physically, the scaling factor $S(Z)$ represents the relativistic binding energy of hydrogen ($Z=1$) to the relativistic binding energy of a hydrogen-like ion with nuclear charge $Z$, both expressed in units of $m_e c^2$. The dependence of the inverse of $S(Z)$ on the nuclear charge $Z$ is visualized in Fig.~\ref{fig:dip1_a0}b, together with its nonrelativistic counterpart. In contrast to the nonrelativistic case, the relativistic function exhibits a steeper increase with increasing $Z$ (see the inset of Fig.~\ref{fig:dip1_a0}b). By incorporating this ratio, the semi-relativistic mapping accounts for the tightening of the atomic structure as $Z$ increases. To adapt these relations to the $\delta$-kick excitation, where $F_0$ scales with the inverse of the time-scaling, Eq.~(\ref{eq.irina}b) is modified to:
\begin{eqnarray}\label{eq.F_ir}
F_0=F'_0\,S^{-1/2}(Z).
\end{eqnarray}

The semirelativistic analogs to Eqs.~\eqref{eq.diP_sig_b} and \eqref{eq.nonsigma}, obtained with the semi-relativistic scaling relations, are:
\begin{subequations}\label{eq.rel_dsig}
\begin{align}
d(t) &= S^{1/2}(Z)\, d'(t'),\\
\sigma(\omega) &= \sigma'(\omega').
\end{align}
\end{subequations}
While other semi-relativistic approaches, such as the field-dressed effective mass method \cite{sfa:tork18,sfa:Vembe2024}, offer excellent agreement between TDSE and TDDE for the hydrogen atom, they typically require the velocity gauge. To maintain consistency with our length-gauge framework and its superior suppression of negative-energy states, we restrict our focus to the scaling methods outlined above.
By systematically comparing these scaling predictions against full TDDE simulations, we can now delineate the regimes where non-relativistic intuition fails and quantify the intrinsic relativistic signatures in the absorption spectra of highly charged ions.

\section{Computational settings} \label{sec:computational}

The calculations presented hereafter are performed in spherical geometry using the B-spline basis framework described in Sec.\ref{sec.rels}. Atomic units (a.u., with $e = m_e = \hbar = 4\pi\varepsilon_0 = 1$ and the $c\approx137$) are adopted for numerical simplicity. To enable a direct comparison across different nuclear charges, all physical observables are presented in scaled coordinates. Time is scaled as $t Z^2$, while the photon energy is represented on a scaled axis defined by $E / Z^2$ and plotted in eV, representing the energy mapped back onto the baseline $Z=1$ hydrogenic scale (where $1\text{ a.u. of energy} \approx 27.2114\text{ eV}$).

For the hydrogen atom ($Z=1$), we employ a basis of order $k=9$ with $N=200$ splines within a radius $R_{\rm max}=100$~a.u. The maximum angular momentum is set to $\ell_{\rm max}=10$, and the time-step is $dt=0.05$~a.u. To maintain a consistent physical resolution of the wavepacket dynamics as the nuclear charge increases, the numerical environment is rescaled for each highly charged ion. Specifically, the box radius and temporal sampling $dt$ are adjusted following the $Z$-scaling laws in Eqs.~(\ref{eq.scal1}a) and (\ref{eq.scal1}b). This ensures that the spatial resolution relative to the hydrogenic Bohr radius $a_0/Z$ and the spectral sampling remain physically equivalent across all species.

To calculate the photoabsorption cross-section, we set a total propagation time of $T = 500$~a.u. for the hydrogenic reference, providing a frequency resolution of $\Delta \omega \approx 2\pi/T$. For high-$Z$ ions, $T$ is scaled by $1/Z^2$ to preserve the number of cycles resolved. While the $1/Z^2$ reduction in time-step significantly increases the computational overhead for heavys ions, this rigorous approach is necessary to isolate the intrinsic relativistic signatures from numerical discretization errors.

\section{Results and Discussion}\label{sec.rels1}

\subsection{Scaling relations for the dipole moment}\label{sec.relsa}
\begin{figure}[h!]
  \hspace*{-0.54cm}
 \includegraphics[width=0.48\textwidth]{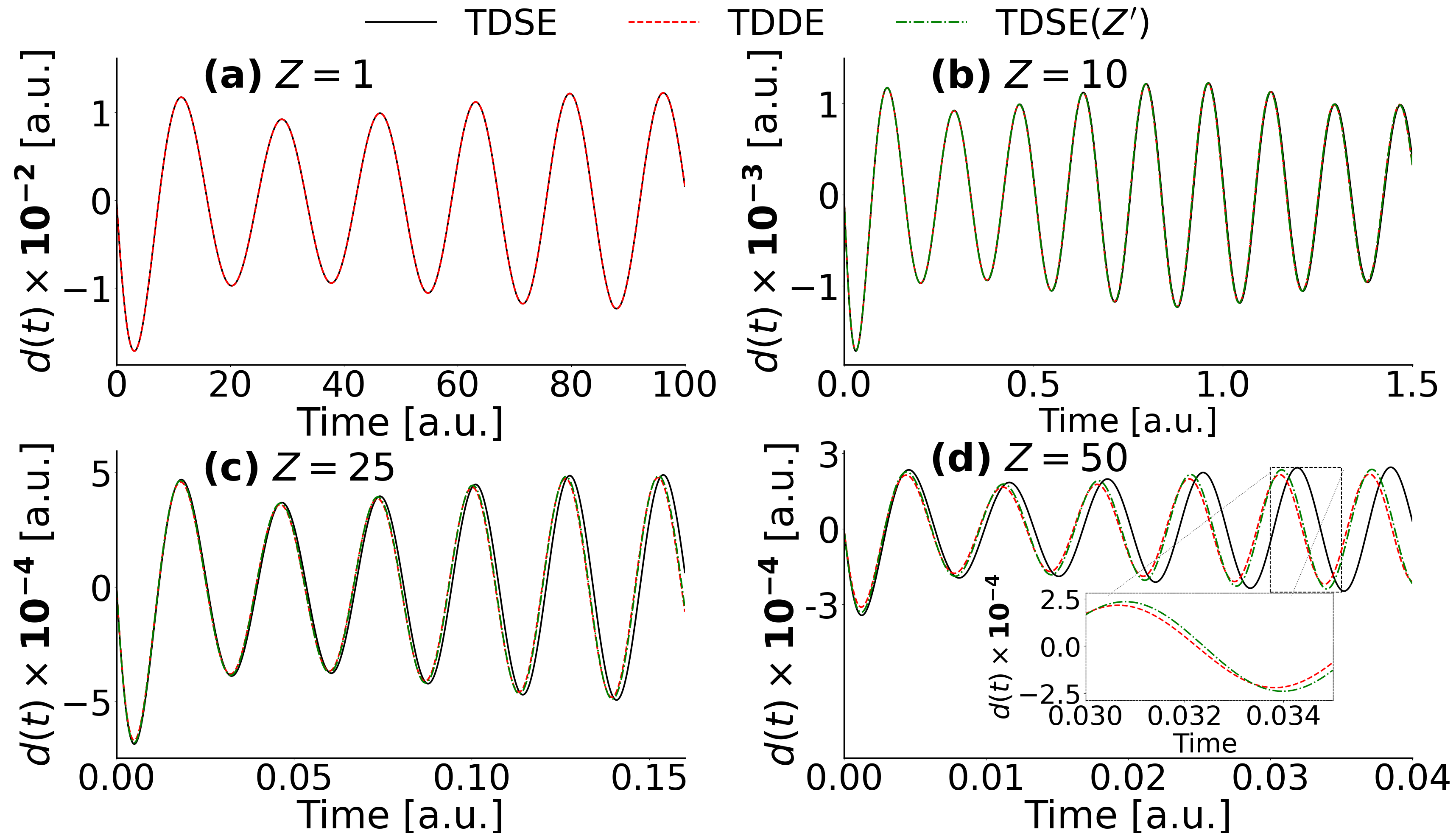}
  \caption{Evolution of the dipole response from the non-relativistic to the relativistic regime. (a) At $Z = 1$ ($F'_0 = 0.01$ a.u.), TDSE and TDDE results coincide. (b) For $Z = 10$ ($F_0 = 0.1$ a.u.), the dynamics accelerate according to the $Z^2$ scaling, with the semi-relativistic TDSE($Z' = 10.007$) tracking the full TDDE result. (c) At $Z = 25$ ($F_0 = 0.25$ a.u.), relativistic influences emerge as a cumulative phase deviation over extended time scales. (d) For $Z = 50$ ($F_0 = 0.5$ a.u.), the relativistic signature becomes dominant, manifesting as a significant temporal compression and frequency up-conversion that the TDSE($Z' = 50.885$) model only partially recovers. }
  \label{fig:dip1}
\end{figure}

We begin by contrasting the relativistic and non-relativistic dipole moments computed from TDSE and TDDE. As expected, the two results are identical for $Z=1$ (Fig.~\ref{fig:dip1}a). This validation confirms that relativistic corrections are negligible for the weak field strengths and low nuclear charges considered here \cite{sfa:milo02a,sfa:reltunKr2002,sfa:tork18,sfa:voitk23}.

For $Z=10$, relativistic and nonrelativistic results remain similar (Fig.~\ref{fig:dip1}b), reflecting the comparable binding energies of the corresponding hydrogen-like ions. The oscillation frequency is substantially higher compared to hydrogen, scaling as $Z^2$ in accordance with Eq.~\eqref{eq.scal1}c. To resolve these accelerated dynamics while maintaining numerical efficiency, the simulation time-step is shortened according to the scaling prescription in Eq.~(\ref{eq.scal1}b). Accordingly, the data are presented over a compressed time interval to resolve individual oscillations. The semi-relativistic TDSE($Z'$) results (green curve), obtained with the effective nuclear charge defined in Eq.\eqref{eq.shz}, closely follow the full TDDE calculations, demonstrating the initial consistency of the energy-matching approach.

As the nuclear charge increases to $Z=25$, relativistic effects become clearly visible over extended time scales (Fig.~\ref{fig:dip1}c), consistent with previous findings for $Z \ge 20$~\cite{gen:gran70,gen:gran07}. For $Z=50$, these effects become dominant (Fig.~\ref{fig:dip1}d), with the relativistic dipole moment exhibiting both an enhanced amplitude and a significant temporal compression compared to its non-relativistic counterpart. These characteristics underscore the dominant role of relativistic dynamics at high $Z$, where increased binding energy and tighter wavepacket confinement fundamentally alter the response to the broadband instantaneous excitation.

For increasingly large $Z$, the TDSE($Z'$) results exhibit a progressive phase deviation from the full TDDE calculations as time advances (Fig.~\ref{fig:dip1}d, inset). This divergence illustrates that while $Z'$-scaling captures the static relativistic correction to the binding energy, it fails to account for the dynamical differences in the transition energy manifold. In the Dirac framework, the energy spacings $\Delta E_{ij}$ do not scale uniformly as they do in the Schr\"odinger case, leading to the observed dephasing in the time-domain dipole. This discrepancy reveals a fundamental limitation of the TDSE($Z'$) framework: while it successfully emulates the ground-state energy, it cannot fully replicate the dynamical phase accumulation dictated by the Dirac equation. This behavior arises because the non-relativistic dipole function $d_{\rm nr}(t)$ in Eq.\eqref{eq.diP_sig_b} and its relativistic counterpart $d_r(t)$ in Eq.\eqref{eq.reldi} involve different sets of energy eigenvalues in their exponential factors, leading to a phase shift that accumulates over the duration of the field-free propagation [Eqs.~\eqref{eq.diP_sig_b} and \eqref{eq.reldi}].

\begin{figure}
  \includegraphics[width=0.48\textwidth]{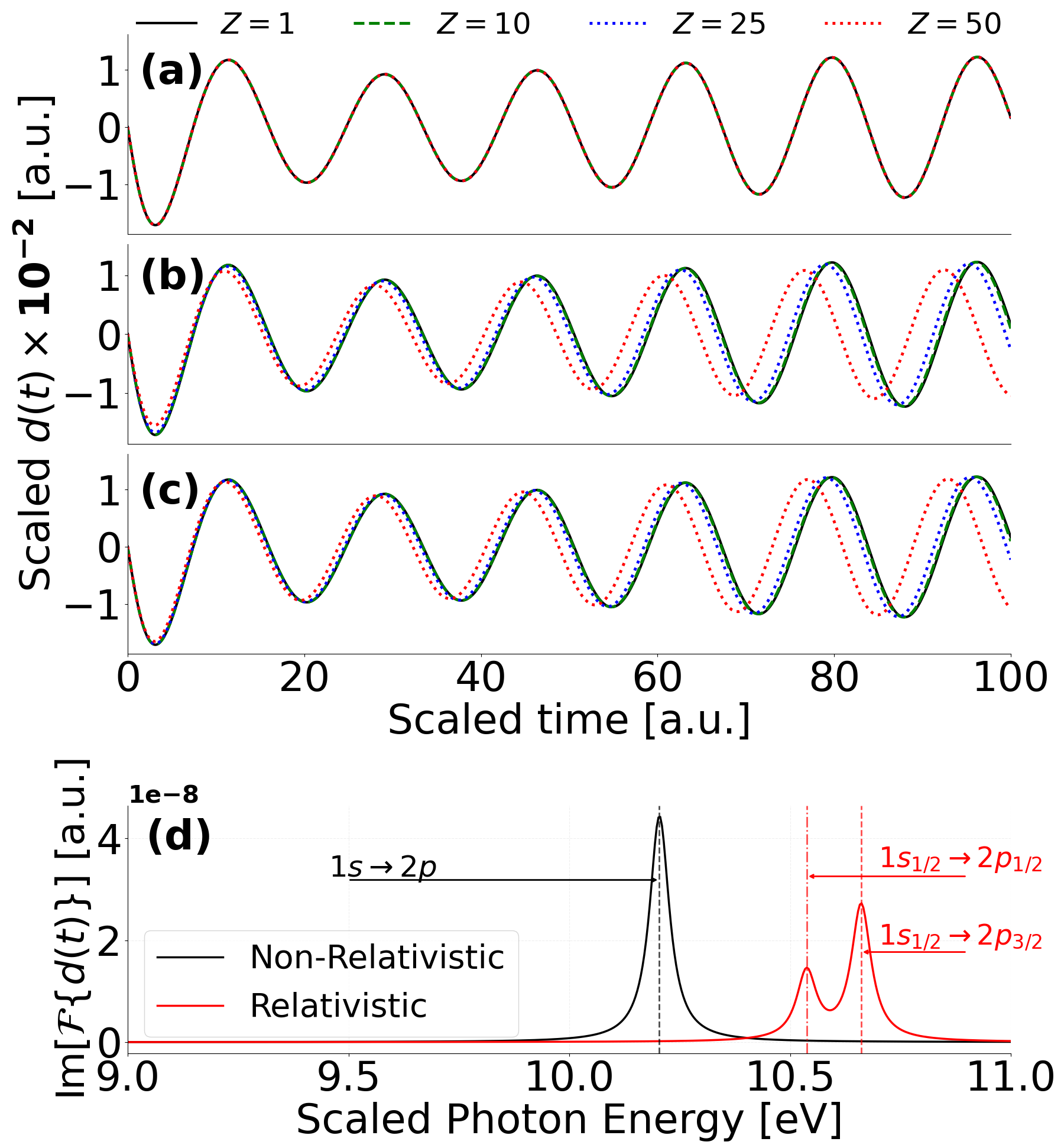} 
 \caption{Temporal evolution of the scaled dipole moment $d(t) \cdot Z$ plotted against scaled time for increasing nuclear charges computed from (a) TDSE, (b) TDDE, and (c) TDSE($Z'$). (d) Fourier transform spectrum of the transient dipole response for $Z=50$ (broadening parameter of 27.2 meV), highlighting the contrast between the non-relativistic singlet from TDSE and the fully relativistic doublet from TDDE. All simulations are performed for a reference field amplitude $F'_0 = 0.01$~a.u.
 }
  \label{fig:hdp_z2_ave}
\end{figure}

The manifestation of these effects depends on the nuclear charge, the field strength, and the duration of the propagation~\cite{sfa:vann12,sfa:ivano18}. These interdependent parameters are unified through the scaling relations, which are validated in Fig.~\ref{fig:hdp_z2_ave}a using the TDSE, with parameters scaled according to Eqs.~\eqref{eq.scal1} and \eqref{eq.scal2}, and the dipole moment adjusted via Eq.~(\ref{eq.sigma_scale}a). The perfect overlap of these curves confirms that non-relativistic hydrogen-like dynamics are entirely predictable from the neutral hydrogen atom benchmark. In contrast, the TDDE results in Fig.~\ref{fig:hdp_z2_ave}b reveal the systematic breakdown of this scaling as $Z$ increases. While the $Z=10$ scaled dipole moment still coincides with the hydrogenic reference, a relevant phase shift emerges at $Z=25$, which becomes the dominant signature at $Z=50$. Notably, the TDSE($Z'$) results in Fig.~\ref{fig:hdp_z2_ave}c successfully recover the essential features of this breakdown. By employing the effective nuclear charge $Z'$, the non-relativistic model aligns more closely with the corresponding relativistic benchmarks, demonstrating the predictive power of the semi-relativistic scaling approach even when the temporal phase tracking is not exact.

In addition to the phase shifts increasing in real time with the atomic number $Z$, the transient polarization contains additional structural information that can be neatly resolved in the frequency domain. As shown in Fig.~\ref{fig:hdp_z2_ave}d for the heaviest considered ion ($Z=50$), the non-relativistic TDSE spectrum features a single resonance corresponding to the degenerate $1s \rightarrow 2p$ transition at the benchmark value of $10.204$~eV (Table~\ref{tab:Z50_transitions}). In contrast, the fully relativistic TDDE dynamics resolve the breaking of this degeneracy into a fine-structure doublet, mapping the $1s_{1/2} \rightarrow 2p_{1/2}$ and $1s_{1/2} \rightarrow 2p_{3/2}$ transitions. The relative intensities of these two components follow the expected branching ratio of  the $(2j+1)$ state degeneracies, confirming that the underlying relativistic angular momentum coupling is accurately preserved in the numerical transition matrix elements. As shown in Table~\ref{tab:Z50_transitions}, the resolved peak positions match the analytical Dirac energy levels with excellent precision, providing an unambiguous frequency-domain signature of the sub-attosecond quantum coherence driven by the broadband impulse.

\begin{table}[H]
\centering
\begin{tabular}{lccc}
\hline
Transition & $E/10^2$ [eV] & $E/25^2$ [eV] & $E/50^2$ [eV] \\
\hline
$1s \rightarrow 2p$ (nr)
& 10.204 & 10.204 & 10.204 \\
$1s_{1/2} \rightarrow 2p_{1/2}$
& 10.217 &10.283 & 10.538 \\
$1s_{1/2} \rightarrow 2p_{3/2}$
& 10.221 & 10.312 & 10.661 \\
\hline
\end{tabular}
\caption{Scaled dipole-allowed relativistic and non-relativistic (nr) transition energies from the ground state to the first excited shell ($n=2$) of hydrogen-like ions with nuclear charges $Z=10$, $Z=25$, and $Z=50$. }
\label{tab:Z50_transitions}
\end{table}

\subsection{Scaling relations for the photo-absorption cross section}\label{sec.relsb}
\begin{figure}[H]
  \centering
  \includegraphics[width=0.5\textwidth]{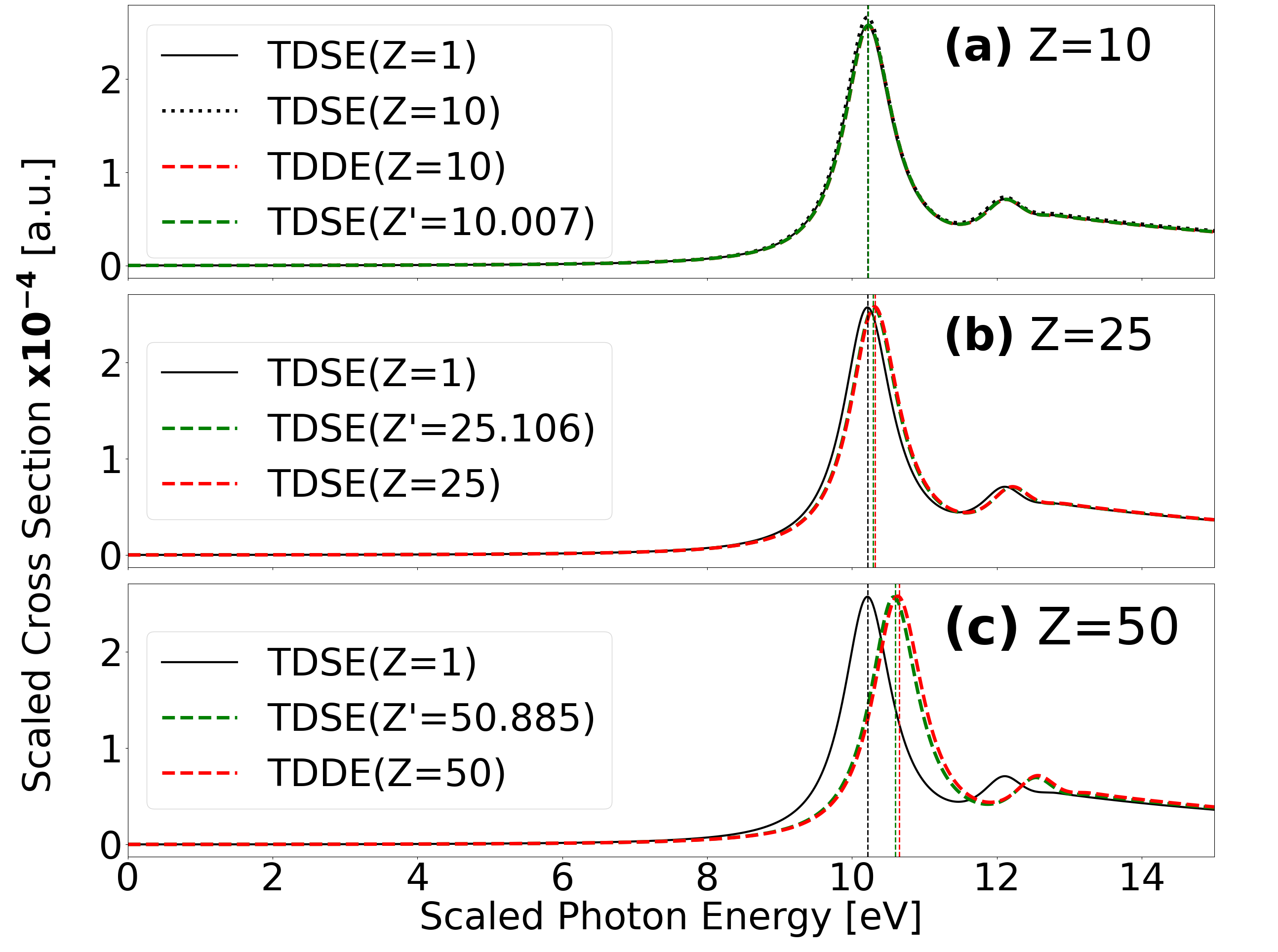}
  \caption{Scaled photo-absorption spectra for hydrogen-like ions with different $Z$, plotted against scaled energy. A broadening parameter of 0.408 eV is used for hydrogen and scaled for the ions via Eq.\eqref{eq.scal1}c (a) $Z =10$, (b) $Z=25$, and (c) $Z=50$ calculated using nonrelativistic and relativistic formulations. The amplitude parameter is fixed at $F'_0 = 0.01$ a.u. for the hydrogen atom ($Z=1$).}
  \label{fig:scaled_sig}
\end{figure}

We complete our analysis by computing the photoabsorption cross-sections.
In the non-relativistic regime, the binding energy of a hydrogen-like ion scales as $Z^2$ with the photoabsorption cross-section following the universal scaling law discussed in Sec.~\ref{sec.theor_1}. This behavior is confirmed in Fig.~\ref{fig:scaled_sig}a, where the scaled hydrogenic benchmark coincides with the hydrogen atom reference (black curve). 
In contrast, the full TDDE solutions (red curve) reveal a systematic relativistic blue shift in the absorption peak. This behavior is expected, as the relativistic energy difference between the ground state and the dipole-allowed first excited state is larger than its nonrelativistic counterpart, as illustrated in Fig.~\ref{fig:dip1_a0}a. This shift, representing the transition from the ground state to the excited manifold (primarily $1s \to 2p$), is negligible at $Z = 10$  (Fig.~\ref{fig:scaled_sig}a) and marginal at $Z=25$ (Fig.~\ref{fig:scaled_sig}b) but becomes pronounced as the nuclear charge rises to $Z=50$ (Fig.~\ref{fig:scaled_sig}c). This behavior represents a clear departure from the non-relativistic benchmark, revealing the spectroscopic signature of the relativistic increase in the central potential depth, which exceeds the $Z^2$ prediction and pushes the absorption resonance toward higher scaled energies. It is worth noting that the fine-structure splitting resolved in the transient dipole spectrum (Fig.~\ref{fig:hdp_z2_ave}d) is not visible in these macroscopic cross-sections. Applying here the same very narrow broadening (27.2 meV) that is needed to spectrally resolve the $j$-dependent components would introduce non-physical spectral oscillations that would prevent a clear identification of the overall relativistic blue shift. To maintain a consistent broadening across the scaled energy axis, the broadening parameter is scaled according to Eq.\eqref{eq.scal1}c.

 Notably, the semi-relativistic TDSE($Z'$) approximation (dashed green curves) fails to reach the quantitative accuracy of the full relativistic treatment. This divergence is rooted in the underlying energy structure: while the non-relativistic transition energy scales quadratically ($\Delta E \propto Z^2$, as shown by the blue curve in Fig.~\ref{fig:dip1_a0}b), the relativistic correction introduces additional $Z$-dependent terms that yield a steeper increase, resulting in the observed spectral shift.

\begin{figure}[H]
 \includegraphics[width=0.5\textwidth]{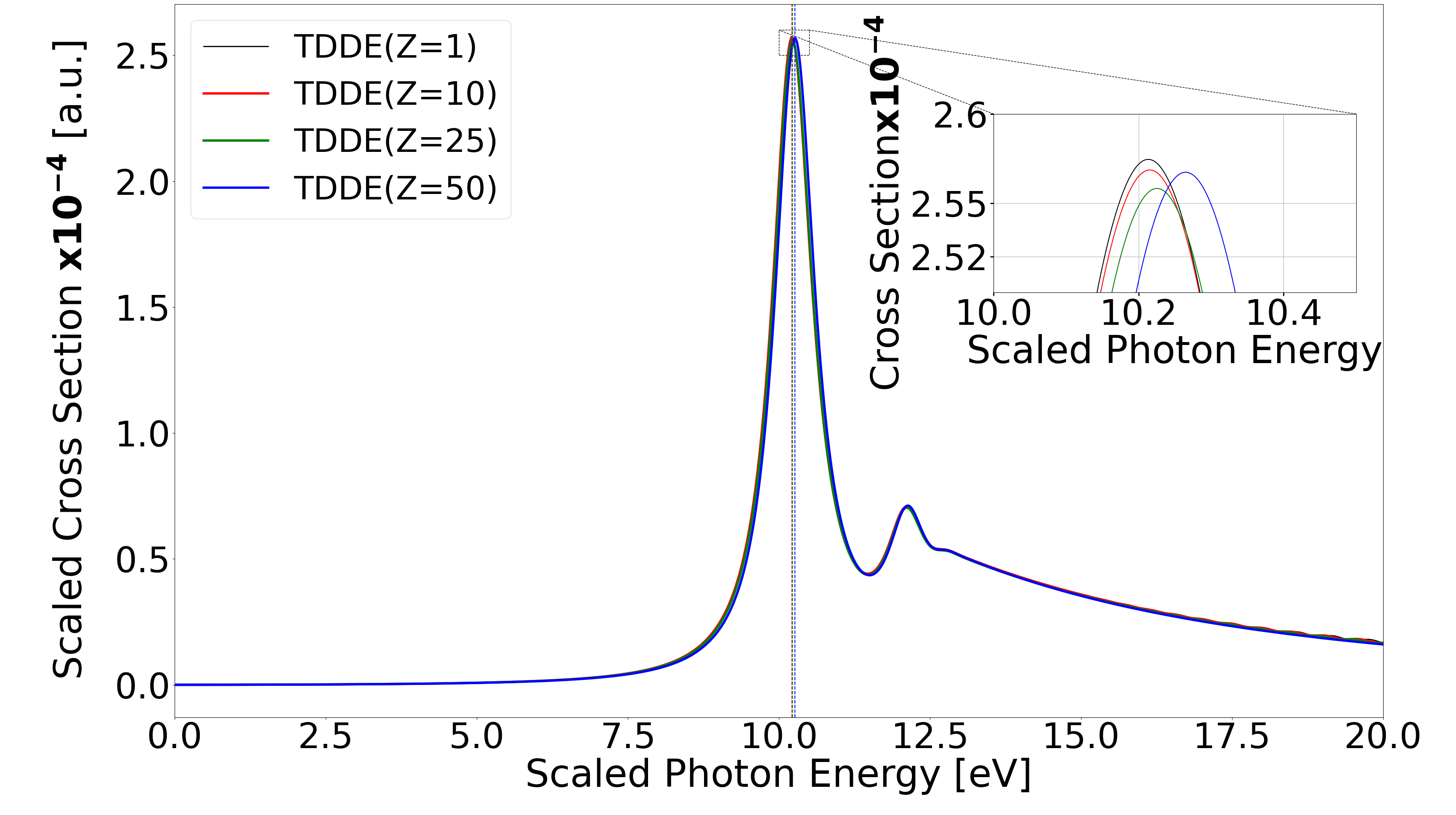} 
  \caption{Scaled photoabsorption cross-section as a function of scaled energy for hydrogen-like ions with various nuclear charges $Z$. The adopted semi-relativistic framework incorporates kinematic corrections that effectively realign the absorption peaks for moderate $Z$. The inset highlights the residual blue shift that reappears for $Z = 50$, marking the transition into the fully relativistic regime where simple frequency mapping is no longer sufficient.  
  The reference amplitude parameter is fixed at $F'_0 = 0.01$~a.u. for the $Z = 1$ case.}
  \label{fig:scaled_sigIr}
\end{figure}

Finally, we test the semi-relativistic scaling relations that incorporate kinematic corrections to the frequency and effectively suppress the blue shift at moderate nuclear charges (Fig.~\ref{fig:scaled_sigIr}). While the overall cross section does not exhibit visible variations across the entire energy scale examined, a close inspection of the peak energy (inset of Fig.~\ref{fig:scaled_sigIr}) reveals a residual blue shift at high $Z$. Furthermore, as demonstrated in Fig.~\ref{fig:dip1_a0}b, the corresponding scaling exhibits a faster increase compared to the nonrelativistic case (see red curve in Fig.~\ref{fig:dip1_a0}b). Consequently, a higher field strength than that expected from conventional nonrelativistic scaling is required. This effectively helps to partially compensate for relativistic effects and yields a more appropriate scaling for the relativistic regime. Nevertheless, this scaling alone is insufficient to fully capture the underlying relativistic dynamics. This behavior indicates that while kinematic frequency mapping accounts for a significant portion of the relativistic effects, it cannot fully reproduce the complex dynamical signatures, such as spin-orbit splitting, the Darwin term, and wavepacket contraction, that become indispensable for an accurate spectral description in the highly relativistic regime.

\section{Summary and Conclusions} 
\label{sec:conclu}
In summary, we investigated the relativistic signatures of hydrogen-like ions interacting with a weak instantaneous broadband excitation. This approach enables the extraction of the full spectroscopic response and the evaluation of the systematic breakdown of non-relativistic scaling symmetries as the nuclear charge $Z$ increases.

The central finding of this work is the emergence of a relativistic blue shift in the photoabsorption cross-section, which becomes relevant for ions with $Z \geq 25$. This spectral shift is the frequency-domain manifestation of a cumulative phase divergence in the time-dependent dipole moment, highlighting the fundamental difference between the TDSE and TDDE frameworks. We demonstrated that the $Z$-scaling relations, while exact in the non-relativistic limit, serve as a rigorous diagnostic tool for quantifying these relativistic kinematics.

Furthermore, we evaluated the predictive limits of semi-relativistic approximations. While the $Z'$-scaling model captures the corrected binding energies, it fails to maintain temporal synchronization with the full TDDE evolution due to the non-uniform scaling of the Dirac energy manifold. However, we have shown that specialized semi-relativistic mappings can effectively suppress the blue shift at moderate $Z$, offering a computationally efficient alternative to the full TDDE. 

Looking forward, our benchmarks not only provide a valuable bridge toward relativistic ion dynamics, defining a clear validity boundary for scaling-based predictions in the emerging field of highly charged ion attosecond spectroscopy. More importantly, they offer an essential framework for interpreting ultrafast light-matter experiments at advanced facilities such as the European XFEL and LCLS-II. As these extreme light sources push toward brighter, sub-femtosecond X-ray pulses, they open up real-time access to the sub-attosecond and zeptosecond dynamics of highly charged ions. In this context, extending our framework to map explicit space-time Dirac probability currents and incorporating many-electron correlations represents exciting avenues for future work, serving as vital steps toward fully deciphering the ultrafast relativistic spectroscopy of complex, highly charged systems.

\section*{Acknowledgments}
\vspace{-2ex}
This work was partially funded by the German Research Foundation (DFG) through the CRC 1375 ``NOA'' (Project number 398816777, subproject A08).

\section*{Data availability}
\vspace{-2ex}

The data that support the findings of this article are openly available in Zenodo:  DOI:10.5281/zenodo.19350553.

\bibliographystyle{apsrev4-2} 
\bibliography{
journals
}

\end{document}